\documentclass{aastex}
\usepackage{emulateapj5}
\usepackage{apjfonts}
\usepackage{epsf}

\slugcomment{UTAP-408}
\shorttitle{SYSTEMATIC EFFECTS ON ARC STATISTICS}
\shortauthors{OGURI}
\begin{document}
\title{Systematic Effects on Tangential and Radial Arc Statistics: 
The Finite Source Size and Ellipticities of the Lens and Source}
%
\author{Masamune Oguri}
\affil{Department of Physics, School of Science, University of Tokyo,
Tokyo 113-0033, Japan.} 
\email{oguri@utap.phys.s.u-tokyo.ac.jp}
%
\received{2002 February 6}
\accepted{2002 March 8}
\begin{abstract}
It has been recognized that the arc statistics of gravitational lensing 
is a useful probe of the density profile of clusters of galaxies. We
examine several systematic effects which are important in predicting 
the number of arcs, with particular attention to the difference between
tangential and  radial arcs. First we derive an analytic expression 
of the cross section for radial arcs taking account of the source size
and find that the moderate source size enhances the cross section for
radial arcs while larger source size ($\gtrsim 1''$ in our example)
 suppresses the number of radial 
arcs. On the other hand, tangential arcs are much less sensitive to the
source size. Next we numerically calculate the cross section for arcs
considering the lens and source ellipticities. We find that the numbers 
of both tangential and radial arcs are highly enhanced by both
ellipticities, by one or two orders of magnitude. The number ratio of
 radial to  tangential arcs is, however, not so affected if the
 threshold axis  ratio of arcs is large ($\gtrsim 7$). The number ratio
 therefore still  remains good statistics which probe the density
 profile of the lens  objects, if the source size effect is correctly
 taken into account.  
\end{abstract} 
%
\section{Introduction}
Clusters of galaxies distort the images of background galaxies due to
the gravitational lensing effect. The statistics of such lensed arcs
have been recognized as a powerful probe of the density profile of lens 
clusters \citep*{wu93,miralda93a,bartelmann96,hattori97,williams99,molikawa99,meneghetti01}. 
In particular, combined statistics of tangential and radial arcs are
useful in determining the density profile of clusters
\citep*{molikawa01,oguri01}. A knowledge of the density profile is
important because of recent indications that the cold dark matter
scenario predicts a cuspy profile \citep*[e.g.,][]{navarro96,navarro97} 
while observations seem to prefer the existence of flat density cores 
\citep*[e.g.,][]{tyson98}.

Most of the previous analytic work assume the infinitesimally small size of
source galaxies. As pointed out by \citet{molikawa01} and \citet{oguri01},
however, the cross section for radial arcs may be severely affected by
the finite source size. Moreover, finite size sources can produce the fold 
image, thus the cross section may be somewhat different from the one 
obtained by assuming the infinitesimal source size. Therefore, we 
analytically calculate the cross section for radial arcs including the
source size effect. We also derive the cross section using Monte Carlo
method and compare this with theoretical predictions.

Another systematic effect is due to the asymmetry in the mass
distribution of the lens cluster
\citep*{bartelmann95a,bartelmann95b,flores00,meneghetti00,meneghetti02}.
\citet{bartelmann95a} claimed that the numerically modeled clusters
produce long arcs about two orders of magnitude more frequently than
spherically symmetric cluster models. Therefore, we also study the
effects of asymmetries of lens objects. Especially we concentrate on the
difference of asymmetry effects between tangential and radial arcs.  As
for the source ellipticity, \citet{keeton01a} analytically calculated
the cross section including the source ellipticity and found that the
effect of the source ellipticity is small compared with the effect of
different density profiles. Nevertheless we also examine the effect of
the source ellipticity in order to check the theoretical predictions and
to see the effect of different threshold axis ratios of arcs.

This paper is organized as follows. In \S \ref{sec:ana}, we analytically 
derive the cross section for radial arcs including the finite source
size. Section \ref{sec:sim} describes the simulation method, and \S
\ref{sec:result} presents the cross section derived by numerical
simulations. The validity of our selection criterion for tangential and
radial arcs is discussed in \S \ref{sec:select}. Finally, we summarize
the results in \S \ref{sec:summary}. Throughout this paper, we assume
the lambda-dominated cosmology $(\Omega_0, \lambda_0)=(0.3,0.7)$.  The
Hubble constant in units of $100{\rm km\,s^{-1}Mpc^{-1}}$ is denoted by
$h$.

\section{Cross Section for Radial Arcs with Finite Source Size}\label{sec:ana}
The previous analytic work often assumed the infinitesimal size of
source galaxies. On the other hand, it has been
pointed out that the finite source size affects the cross section
particularly for radial arcs \citep{molikawa01,oguri01}. Moreover,
sources with a finite size have a possibility to produce the fold image
which is formed by merging two images near the radial critical curve,
even in the case of the spherical symmetric lens. Therefore, the cross
section for radial arcs including the finite source size effect may be
somewhat different from the one calculated by neglecting the source size.
In this section, we derive an analytic expression of the cross section
for radial arcs taking account of the finite size of source galaxies
assuming spherical lenses.

The image position $\vec{\xi}$ in the lens plane corresponding to the
source at $\vec{\eta}$ in the source plane is determined by the lens
equation \citep*[e.g.,][]{schneider92}:
\begin{equation}
 y=x-\alpha(x),
\end{equation}
where $x=|\vec{x}|=|\vec{\xi}|/\xi_0$,
$y=|\vec{y}|=|\vec{\eta}|D_{\rm OL}/(\xi_0D_{\rm OS})$, 
and $D_{\rm OL}$ and $D_{\rm OS}$ denote the angular
diameter distances from the observer to the lens and the source planes,
respectively. The normalization length $\xi_0$ can be arbitrary.
The deflection angle $\alpha(x)$ is related to the mass
distribution of the lens object. In the case of spherical lenses, a
source at $x$ is stretched by the factor $\mu_{\rm
t}(x)\equiv(y/x)^{-1}$ along the tangential direction and $\mu_{\rm
r}(x)\equiv(dy/dx)^{-1}$ along the radial direction. Then for the
infinitesimal source, the cross sections for tangential and radial arcs
are simply given by the areas in the source plane in which the
inequalities:
\begin{eqnarray}
 \mbox{Tangential arc}&:\hspace{1mm}&T(x)\equiv\left|\frac{\mu_{\rm t}(x)}{\mu_{\rm r}(x)}\right|\geq\epsilon_{\rm th}\label{tth}\\
 \mbox{Radial arc}&:\hspace{1mm}&R(x)\equiv\left|\frac{\mu_{\rm r}(x)}{\mu_{\rm t}(x)}\right|\geq\epsilon_{\rm th}\label{rth}
\end{eqnarray}
are satisfied. For the threshold axis ratio, e.g., \citet{oguri01}
adopted $\epsilon_{\rm th}=4$. Since the assumption that the source size
is sufficiently small seems to be valid for usual tangential arcs
\citep[see][]{hattori97}, in this paper we use equation (\ref{tth}) to
predict the cross section for tangential arcs.

\vspace{0.5cm}
\centerline{{\vbox{\epsfxsize=7.5cm\epsfbox{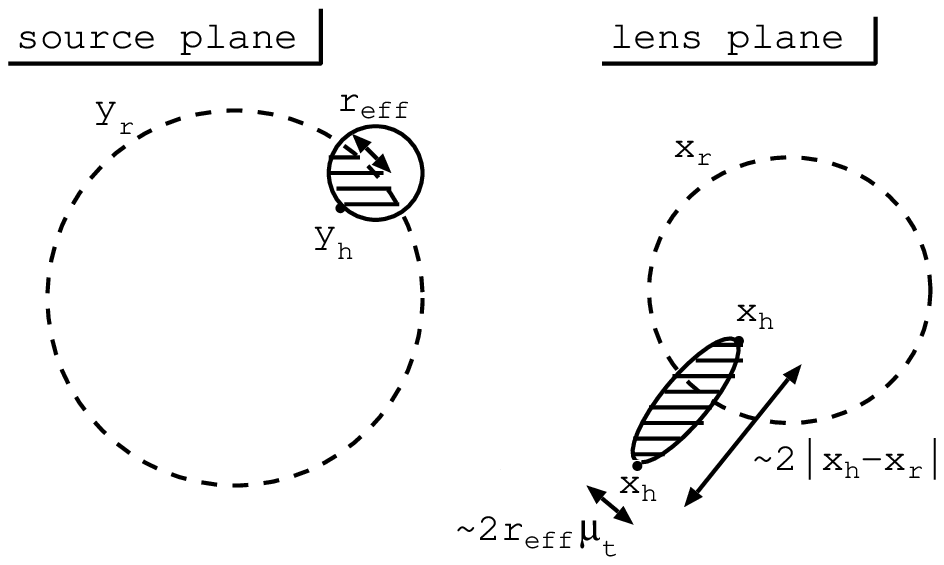}}}}
\figcaption{A schematic diagram of the lens mapping in forming a fold
 image. The radial critical line (with radius $x_{\rm r}$) and radial
 caustic (with radius $y_{\rm r}$) are shown by dashed lines. If the
 source cross the radial caustic, only a part of the source inside the
 radial caustic (denoted by the shading) is mapped into near the radial
 critical line. Then $r_{\rm eff}$ (eq. [\ref{reff}]) means the
 effective radius of the shaded part of the source. The radius of the
 point in the source which is nearest from the center of the lens is
 denoted by $y_{\rm h}$, and either radius of corresponding two points
 in the lens plane is $x_{\rm h}$. In this situation, the criterion for
 radial arcs is well approximated by equation (\ref{rf}). \label{fig:radial}}  
\vspace{0.5cm}

Consider the cross section for radial arcs when the source has the
finite size. First we consider the fold images. These images are
produced only when sources cross the radial caustic:
\begin{equation}
 y_{\rm r}-r_{\rm S}\equiv y_-\leq y\leq y_+\equiv y_{\rm r}+r_{\rm S},
 \label{fold}
\end{equation}
where $y_{\rm r}$ is the radius of the radial caustic and $r_{\rm S}$ is the
dimensionless source radius. In this situation, we approximate the
criterion for radial arcs as (see Figure \ref{fig:radial})
\begin{equation}
 R_{\rm f}(x_{\rm h})\equiv\frac{\left|x_{\rm h}-x_{\rm r}\right|}{\mu_{\rm t}(x_{\rm r})r_{\rm eff}}\geq\epsilon_{\rm th},
\label{rf}
\end{equation}
where $x_{\rm r}$ is the radius of the radial critical line and $x_{\rm
h}$ denotes two farthest points from the radial critical line. The
latter is related to the center of the source $y$ as
\begin{equation}
 y=y_{\rm h}-r_{\rm S},
\end{equation}
with $y_{\rm h}=|y(x_{\rm h})|$ being the radius corresponding to
$x_{\rm h}$. The effective radius of the source $r_{\rm eff}$ in
equation (\ref{rf}) is defined by
\begin{eqnarray}
  r_{\rm eff} \equiv \left\{
      \begin{array}{ll}
        \displaystyle{\sqrt{r_{\rm S}^2-(y_{\rm r}+r_{\rm S}-y_{\rm h})^2}}&
        (y_{\rm r}<y_{\rm h}-r_{\rm S}), \\ 
        \displaystyle{r_{\rm S}}&
        (y_{\rm r}\geq y_{\rm h}-r_{\rm S}).
      \end{array}
   \right. \label{reff}
\end{eqnarray}
We calculate the cross section as the area in the source plane in which
the condition (\ref{rf}) is satisfied at both sides of the radial critical
line in the image plane, that is, the area in which the following
condition is satisfied:
\begin{equation}
 y<y_{\rm f}\equiv \min(y_{{\rm h}+}-r_{\rm S},y_{{\rm h}-}-r_{\rm S}),
 \label{foldlim}
\end{equation}
where $y_{{\rm h}\pm}=|y(x_{{\rm h}\pm})|$ are two radii corresponding to 
solutions of the equation $R_{\rm f}(x_{{\rm h}\pm})=\epsilon_{\rm th}$ at
both sides of the radial critical line.

Next consider images which do not touch the radial critical line. These
images are possible only for
\begin{equation}
y<y_-.
\label{sep}
\end{equation}
Except for this condition, the cross
section can be calculated similarly as the one with infinitesimal
source size, and is explicitly expressed as
\begin{equation}
 y_{\rm r}>y>y_{\rm u}\equiv\min(y_{{\rm u}+},y_{{\rm u}-}),
 \label{notfoldlim}
\end{equation}
where $y_{{\rm u}\pm}=|y(x_{{\rm u}\pm})|$ are two radii corresponding to
solutions of the equation $R(x_{{\rm u}\pm})=\epsilon_{\rm th}$ at
both sides of the radial critical line.

Combining equations (\ref{fold}), (\ref{foldlim}), (\ref{sep}), and
(\ref{notfoldlim}), the cross section for radial arcs $\sigma_{\rm rad}$
is written as 
\begin{equation}
 \sigma_{\rm rad}=\left(\frac{\xi_0D_{\rm OS}}{D_{\rm OL}}\right)^2S(\min(y_+,y_f),\max(y_-,y_u)),\label{cs_rad}
\end{equation}
where $S(a,b)$ is defined by
\begin{eqnarray}
 S(a,b)\equiv \left\{
      \begin{array}{ll}
        \displaystyle{\pi(a^2-b^2)}&
        (a>b), \\ 
        \displaystyle{0}&
        (a<b).
      \end{array}
   \right. 
\end{eqnarray}
The radii $y_+$, $y_-$, $y_{\rm f}$, and $y_{\rm u}$ are defined by
equations (\ref{fold}), (\ref{fold}), (\ref{foldlim}), and
(\ref{notfoldlim}), respectively. 

\vspace{0.5cm}
\centerline{{\vbox{\epsfxsize=7.5cm\epsfbox{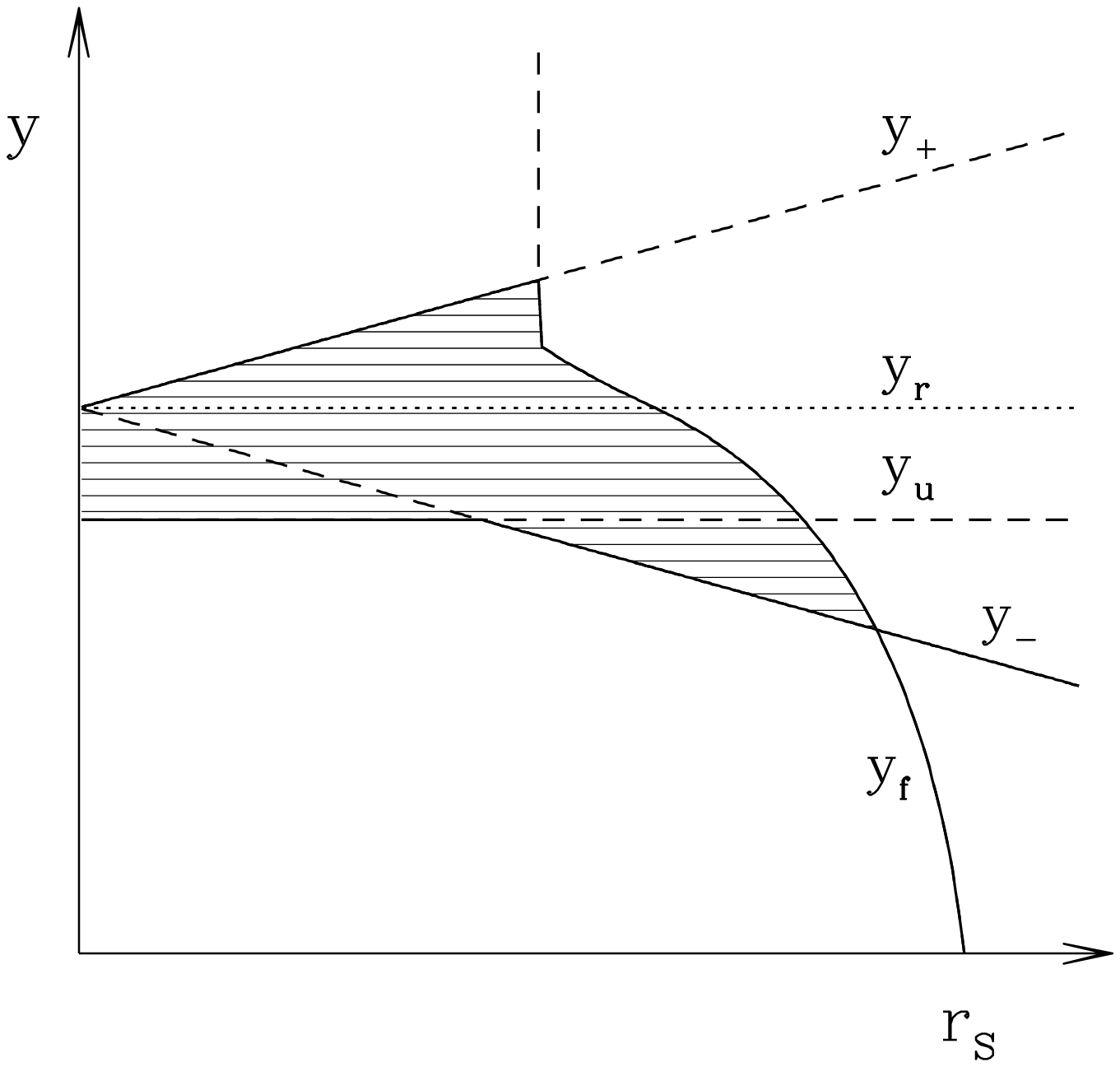}}}}
\figcaption{Dependence of radii defined in order to calculate the cross
 section for radial arcs, $y_+$ (eq. [\ref{fold}]), $y_-$ (eq.
 [\ref{fold}]), $y_{\rm f}$ (eq. [\ref{foldlim}]), and 
 $y_{\rm u}$ (eq. [\ref{notfoldlim}]), on the source radius $r_{\rm S}$.
 The radius of the radial caustic $y_{\rm r}$ is also shown for
 reference. The definition of each radius is given in \S \ref{sec:ana}.
 The shading indicates the region where radial arcs are
 formed.\label{fig:plotline}}
\vspace{0.5cm}

Figure \ref{fig:plotline} shows the dependence of each radius defined in
the above on the source radius $r_{\rm S}$. This figure indicates
that the moderate amount of source size {\it enhances} the cross
section, while the larger source size significantly decreases the cross
section. 

\section{Simulation Method}\label{sec:sim}

In this section, we briefly describe the mapping and detecting algorithm of
arcs. Most of our method described in \S \ref{sec:map} and \S
\ref{sec:rec} follows the work by \citet{miralda93b} and
\citet{bartelmann94}. 

\subsection{Lens Model}
For density profiles of clusters, we adopt the generalization of those
proposed by \citet{navarro96,navarro97}, the generalized NFW profile
\citep{jing00}:  
\begin{equation}
 \rho(r)=\frac{\rho_{\rm crit}\delta_{\rm c}}
{\left(r/r_{\rm s}\right)^\alpha\left(1+r/r_{\rm s}\right)^{3-\alpha}}.
\label{generalizednfw}
\end{equation}
For this profile, we choose the normalization of the lens equation as
$\xi_0=r_{\rm s}$.  
The concentration parameter $c_{\rm vir}=r_{\rm vir}/r_{\rm s}$, where $r_{\rm
vir}$ is the virial radius of dark halo, depends on the halo mass $M$ and 
the redshift $z_{\rm L}$. We calculate $c_{\rm vir}$ using the fitting
form derived by \citet{bullock01}:
\begin{equation}
 c_{\rm vir}=\frac{8}{1+z_{\rm L}}\left(\frac{M}{10^{14}h^{-1}M_\odot}\right)^{-0.13},
\end{equation}
for $\alpha=1$, and we generalize it to $\alpha\neq 1$ by multiplying
$(2-\alpha)$ \citep[see][]{keeton01b}. 

Ellipticities of the lens and source are included as follows. For the
lens ellipticity, we simply substitute $u$ for $x$ in the axially
symmetric lens potential $\psi(x)$:
\begin{equation}
 u^2=(1-e_{\rm L})x_1^2+\frac{x_2^2}{1-e_{\rm L}},
\end{equation}
where $\vec{x}=(x_1,x_2)$. With this substitution, the ellipticity is
$e_{\rm L}=1-b/a$, where $a$ and $b$ are the semi-major and semi-minor
axes. Then the deflection angle $\alpha$ is obtained by
\begin{eqnarray}
 \alpha(x_1)&=&\frac{(1-e_{\rm L})x_1}{u}\alpha(u),\label{alpha1}\\
 \alpha(x_2)&=&\frac{x_2}{(1-e_{\rm L})u}\alpha(u),\label{alpha2}
\end{eqnarray}
where $\alpha(u)$ is the deflection angle for the axially symmetric
lens. Note that this elliptical model is same as the one adopted by
\citeauthor{meneghetti02} (\citeyear{meneghetti02}; see also
\citealt{golse02}). The source ellipticity is included in the 
detection of images (eq. [\ref{withinsource}]). 

\subsection{Lens Mapping}\label{sec:map}
To begin with, we choose a sufficiently large region ($\sim
3'\times3'$) in the lens plane in which all the arcs exist. In this
region we prepare 
regular grids. Each grid point is denoted by $(x_{1i}, x_{2j})$, where
integers $i$ and $j$ are restricted in $1\leq i, j\leq N_{\rm grid}$. In
the practical calculations, we adopt $N_{\rm grid}=8192$ throughout the
paper. Given the deflection angle $\vec{\alpha}$, for all grid
points we can calculate the source point $(y_1(i,j), y_2(i,j))$ which
corresponds to $(x_{1i}, x_{2j})$ by using the lens equation. 

Next we consider the source with center $(y_{\rm 1c}, y_{\rm 2c})$ and
dimensionless radius $r_{\rm S}$ and ellipticity $e_{\rm S}$. We
regard the grid point $(x_{1i}, x_{2j})$ is a part of lensed images if
the following condition is satisfied:  
\begin{equation}
 \frac{\left[y_1(i, j)-y_{\rm 1c}\right]^2}{r_{\rm S}^2/(1-e_{\rm S})}+\frac{\left[y_2(i, j)-y_{\rm 2c}\right]^2}{r_{\rm S}^2(1-e_{\rm S})}\leq 1.
\label{withinsource}
\end{equation}
For all the grids we check this condition and obtain the pattern of
lensed images. In general, multiple images can be generated by
gravitational lensing. Thus we search each image grid and recognize
neighboring image grids as the same image. The magnification of the
image is then proportional to the number of grid points it contains.

\subsection{Detection of Arcs}\label{sec:rec}
To analyze the lens properties, we calculate the magnification $\mu$,
length $l$, width $w$, orientation $\phi$ of the image. For the multiply
lensed system, we calculate these quantities for each image. 

First, the magnification is easily calculated from the number of grid
points $N_{\rm image}$ which are recognized as the image; 
$\mu=N_{\rm image}(\Delta x)^2/\pi r_{\rm S}^2$, where $\Delta x$ is the 
dimensionless interval of grids. This gives fairly accurate 
values because an image contains many grid points in our calculations; 
typically $N_{\rm image}\sim 600$. Next we calculate the length as follows.
First we search the center of the image C as the point at which the value 
of the left hand side in equation (\ref{withinsource}) becomes smallest. 
Then we find the point A in the image which is farthest away from the center. 
Then we find the point B, also along the image, which is farthest away from 
the point A. We calculate the length of the image by 
$l={\rm \overline{AC}+\overline{BC}}$. The width is simply taken such 
that $\pi lw=\mu\pi r_{\rm S}^2$. We define the orientation of the image 
as the angle between the normal of the segment joining A and B and the
segment joining the origin and the center of the image. The orientation
$\phi$ takes the values in the range of $0^\circ\leq\phi\leq90^\circ$. 

Using the above quantities, we define the tangential and radial arcs as
\begin{eqnarray}
 \mbox{Tangential arc}&:\hspace{1mm}&\frac{l}{w}\geq\epsilon_{\rm th}\;\;\mbox{and}\;\;\phi\leq40^\circ,\label{detecttan}\\
 \mbox{Radial arc}&:\hspace{1mm}&\frac{l}{w}\geq\epsilon_{\rm th}\;\;\mbox{and}\;\;\phi\geq50^\circ,\label{detectrad}
\end{eqnarray}
where $\epsilon_{\rm th}$ is the threshold axis ratio.
The validity of this selection criterion is discussed in \S
\ref{sec:select}.

\tabcaption{The canonical parameter set used in this
 paper\label{table:arcmap_cond}}

\begin{center}
\begin{tabular}{ll}
\tableline\tableline\noalign{\smallskip}
 Parameters & Values \\
\noalign{\smallskip}
\tableline\noalign{\smallskip}
Number of grids & $8192^2$ \\
  Lens mass       & $10^{15}h^{-1}M_\odot$ \\
  Lens redshift   & 0.3 \\
  Source redshift & 1.2 \\
  Source diameter & $0.5^{''}$ \\
\noalign{\smallskip}
\tableline
\noalign{\smallskip}

\end{tabular}
\end{center}

A canonical parameter set used in this paper is summarized in Table
\ref{table:arcmap_cond}. With this number of grids, a single mapped image
contains typically $\sim 600$ grids, so it is sufficient to resolve the 
shape of images. We examine the shapes of the images of
$50000\sim200000$ sources and then estimate the cross sections. We also
estimate errorbars of numerically calculated cross sections simply by
statistical errors.

\section{Numerical Calculation of Cross Sections}\label{sec:result}

\begin{figure*}[tb]
\begin{center}
\leavevmode
\epsfxsize=13.0cm 
\epsfbox{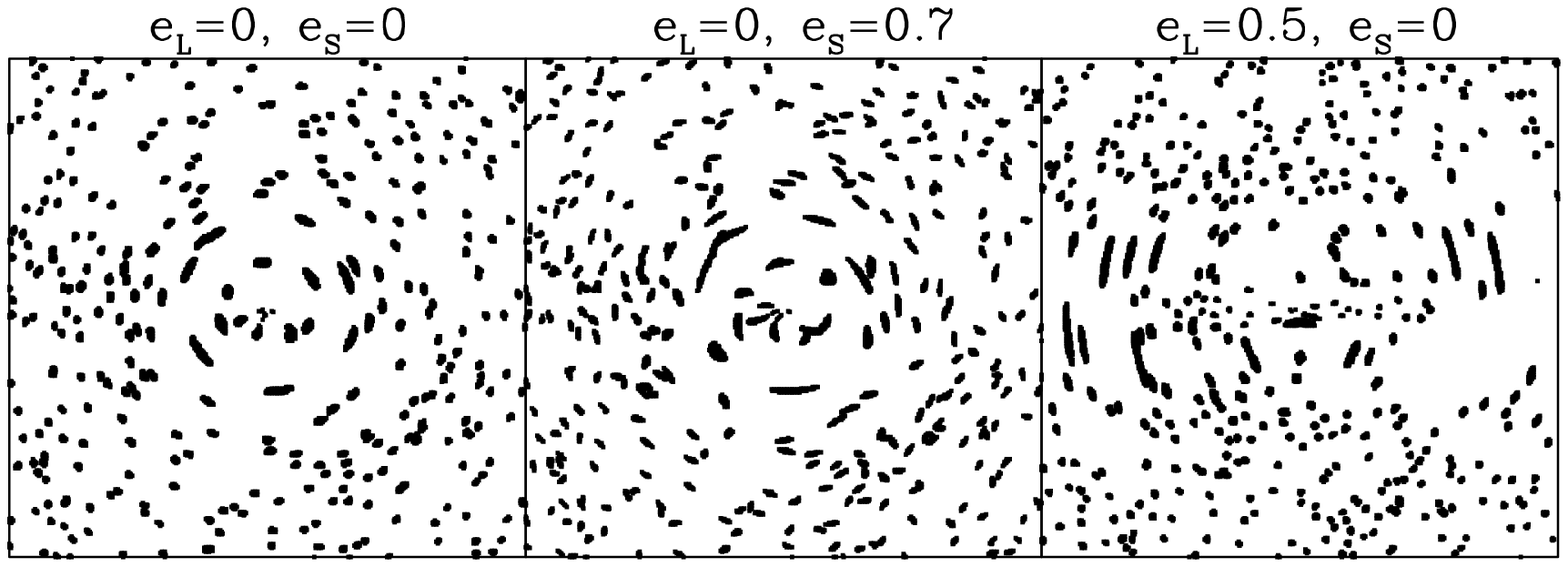}
\caption{Snapshots of simulated images with and without ellipticities,
 produced from the same spatial distribution of source galaxies. The
 source ellipticity is denoted by $e_{\rm S}$ (eq. [\ref{withinsource}])
 while the lens ellipticity is $e_{\rm L}$ (eqs. [\ref{alpha1}] and
 [\ref{alpha2}]).\label{fig:map_mono}}
\end{center}
\end{figure*}

We numerically calculate the cross sections for tangential and radial
arcs using the method described in \S \ref{sec:sim}. First we show
examples of simulated images with and without ellipticities in Figure
\ref{fig:map_mono}. These plots clearly show that the effects of
ellipticities are indeed large and cannot be neglected. Therefore, in
this section, we numerically study the effects of source and lens
ellipticities as well as the finite source size effect.

\begin{figure*}[tb]
\begin{center}
\leavevmode
\epsfxsize=10cm 
\epsfbox{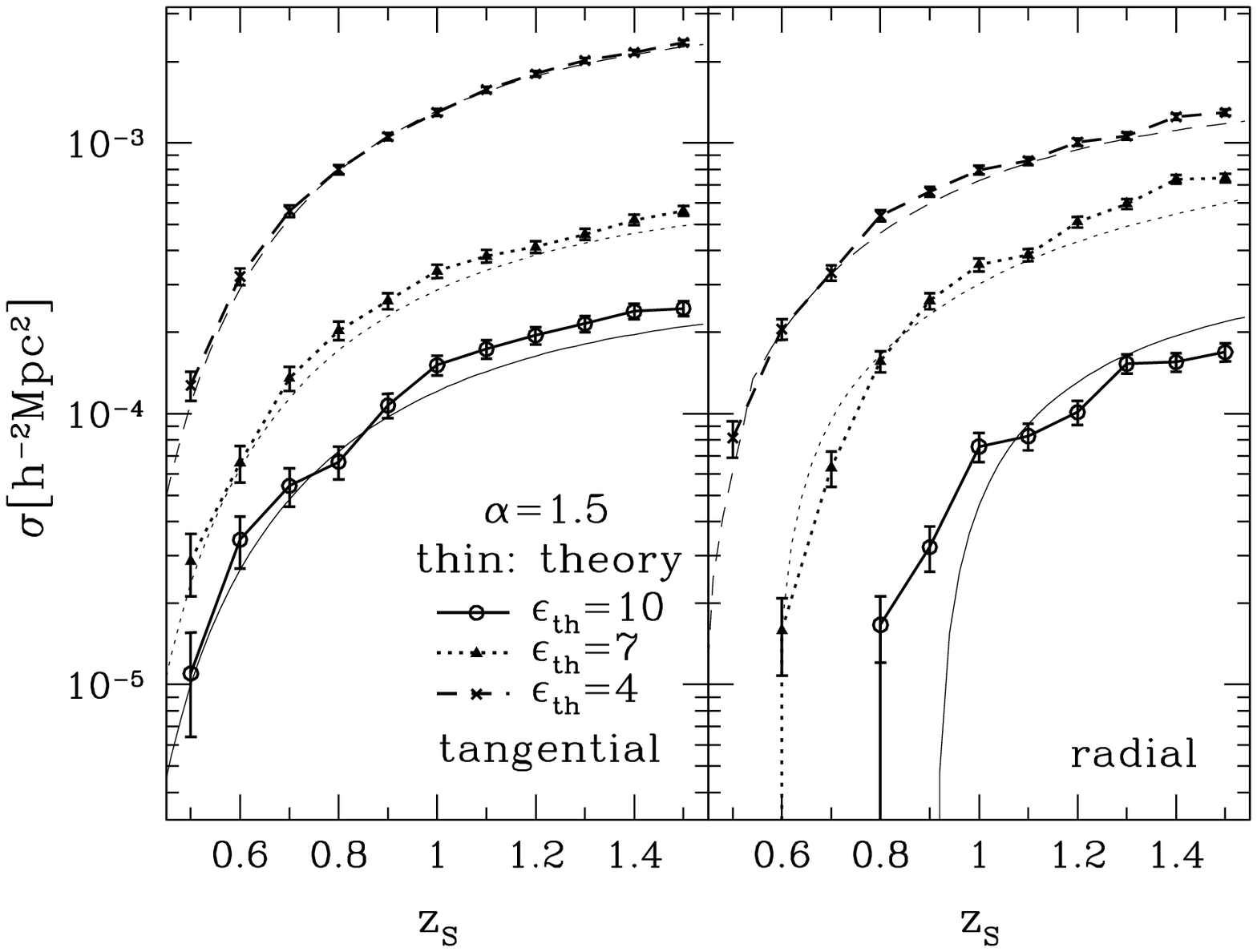} 
\caption{The lensing cross sections for tangential ({\it left panel})
 and radial ({\it right panel}) arcs, derived from numerical
 simulations, against the source redshift  
 $z_{\rm S}$. Ellipticities are not included. The density profile is
 the generalized NFW profile (eq. [\ref{generalizednfw}]) with 
$\alpha=1.5$. For the detection of
 arcs, three criteria of axis ratio are adopted; $\epsilon_{\rm th}=10$
 ({\it solid}), $7$ ({\it dotted}), and $4$ ({\it dashed}). The
 errorbars indicate statistical errors in estimating the cross sections
 numerically. Thin lines  are theoretical predictions which are
calculated from equation (\ref{tth}) for tangential arcs and 
equation (\ref{cs_rad}) for radial arcs.\label{fig:zs}}
\end{center}
\end{figure*}

In Figure \ref{fig:zs}, to check simulated cross
sections, we plot the lensing cross sections for tangential and 
radial arcs against the source redshift $z_{\rm S}$, in the case of no
ellipticities. The density profile is the generalized NFW profile (eq.
[{\ref{generalizednfw}}]) with $\alpha=1.5$. In what follows, we
consider three criteria of axis ratio for the detection of arcs;
$\epsilon_{\rm th}=10$, $7$, and $4$. Theoretical predictions, which are
calculated from equation (\ref{tth}) for tangential arcs and 
equation (\ref{cs_rad}) for radial arcs, are also plotted in thin lines.
These plots make sure that the cross sections for tangential and radial
arcs show good agreement between theoretical predictions and
simulations.

\subsection{Finite Source Size}\label{sec:finitesource}

\begin{figure*}
\begin{center}
\leavevmode
\epsfxsize=10cm 
\epsfbox{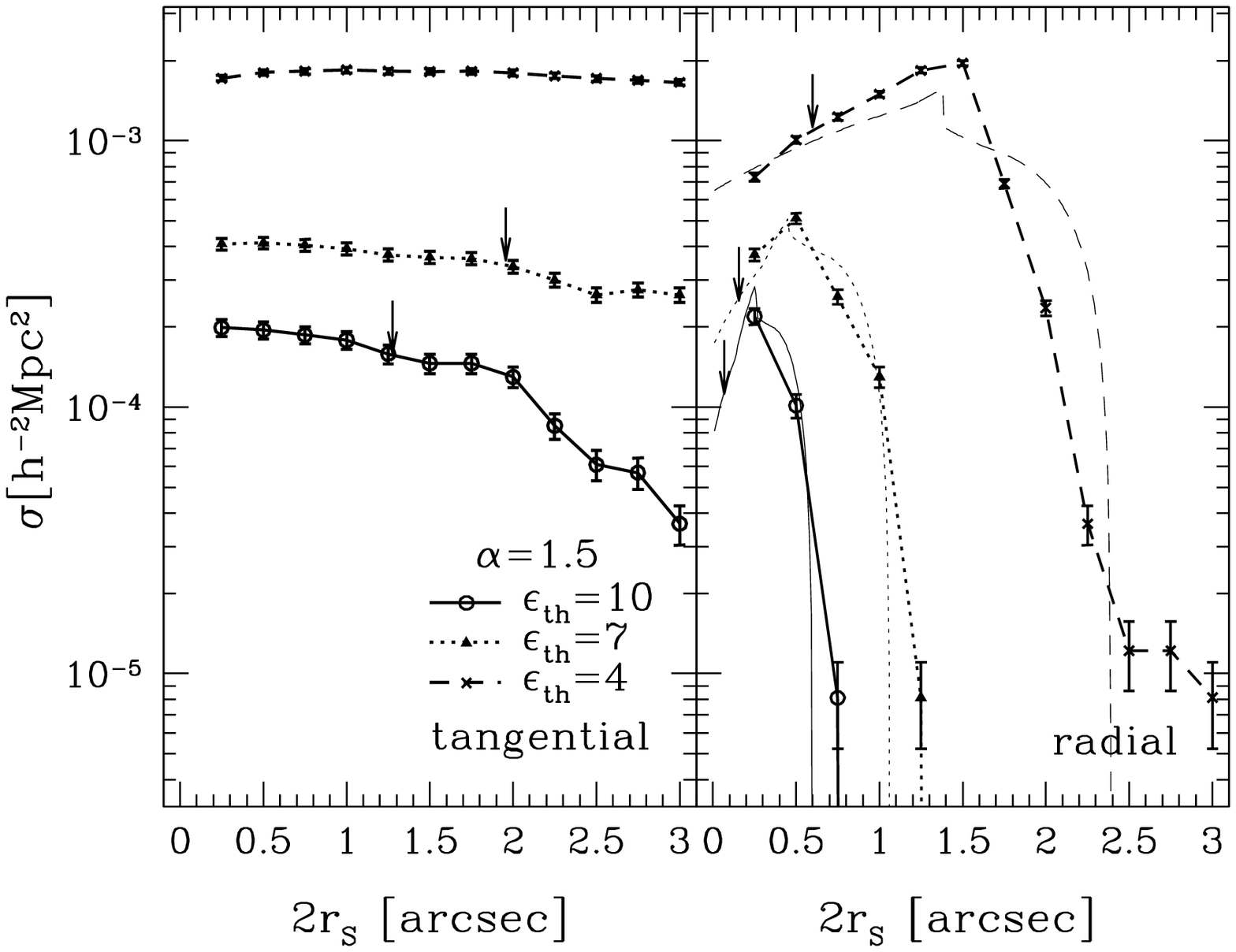} 
\caption{The effect of finite source size. The lensing cross sections
 for tangential ({\it left panel}) and radial ({\it right panel}) arcs
 with $\alpha=1.5$ generalized NFW profile (eq. [\ref{generalizednfw}])
 are plotted against the source diameter $2r_{\rm S}$ in units of arcsecond. 
 Theoretical predictions of the cross section for radial arcs 
 (eq. [\ref{cs_rad}]) are shown by thin lines.
 Arrows indicate the cutoff size of cross sections
 adopted in \citet{oguri01}. Note that the typical
 angular diameter of sources (faint galaxies, $B\gtrsim25$) is $\sim1^{''}$
 \citep[e.g.,][]{lilly91}.\label{fig:rs}}
\end{center}
\end{figure*}

In Figure \ref{fig:rs}, we show the effect of the finite source size.
We also plot the theoretical prediction of cross sections for radial arcs
(eq. [\ref{cs_rad}]) and find that our analytic calculations show fairly
good agreement with simulations.
These plots clearly indicate that the source size severely affects the
cross section for radial arcs. The longer arcs (larger $\epsilon_{\rm
th}$) are more strongly affected by the source size than shorter arcs.
Therefore, we need to know the source size and properly take account of
the finite source size effect in calculating the number of radial arcs.
On the other hand, the number of tangential arcs does not change so much
within the range we examined. The typical
 angular diameter of sources (faint galaxies, $B\gtrsim25$) is $\sim1^{''}$
 \citep[e.g.,][]{lilly91}, and this is small compared with the cross
 sectional region for tangential arcs. This implies that we can estimate the
lower limit of inner slope of density profile, $\alpha$, from
observations of radial arcs if we use the maximum cross section for radial 
arcs, because the density profiles with larger $\alpha$ usually produce the 
larger number of radial arcs. We also found that the size of cutoff of 
cross sections adopted in \citet{oguri01} is not so adequate, which
corresponds to the width of the cross sectional region and is described 
by arrows in Figure \ref{fig:rs}; the cross 
sections begin to decrease when the source diameter ($2r_{\rm S}$) is two 
times the cutoff width. From this figure, the cross
section for radial arcs have a steep cutoff against the source size
while the cross section for tangential arcs seems to decrease mildly as
the source becomes larger. We conclude that the radial arcs are more
sensitive to the source size effect also from this point.

\subsection{Source Ellipticity}

\begin{figure*}
\begin{center}
\leavevmode
\epsfxsize=10cm 
\epsfbox{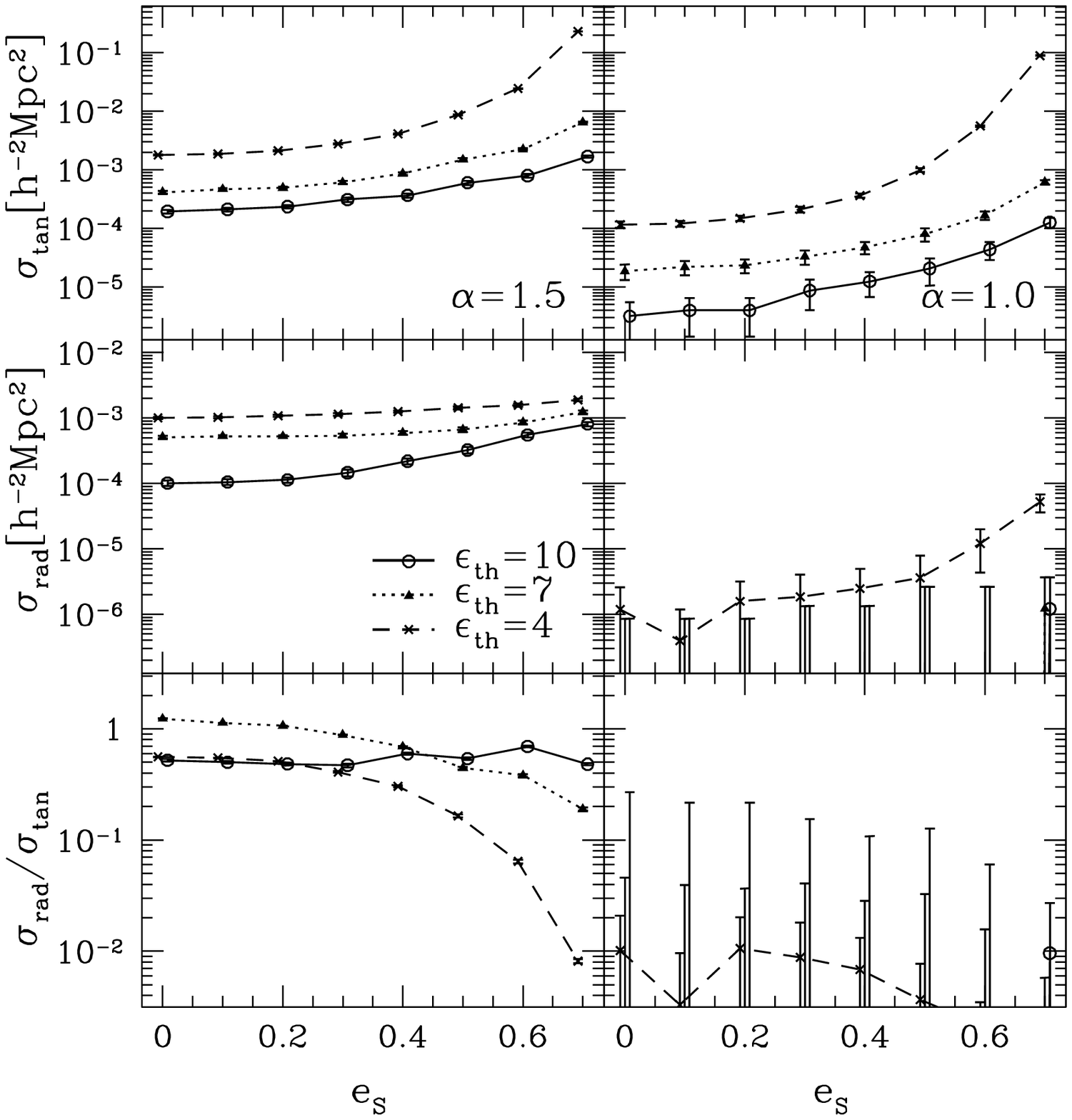} 
\caption{The effect of the source ellipticity. For the density profile of
 the lens object, we consider the generalized NFW  (eq.
 [\ref{generalizednfw}]) with $\alpha=1.5$ ({\it 
 left panels}) and $\alpha=1.0$ ({\it right panels}). From top to
 bottom, we plot the cross sections for tangential arcs, radial arcs,
 and the ratio of radial to tangential arcs.\label{fig:sel}}
\end{center}
\end{figure*}

Figure \ref{fig:sel} shows the effect of the source ellipticity. As
pointed out by \citet{keeton01a}, the source ellipticity increases the
number of both tangential and radial arcs, but decreases the
number ratio of radial to tangential arcs. Although \citet{keeton01a}
analytically derived the cross section including the source ellipticity
assuming the infinitesimal source size and showed the above dependence, we 
confirmed this by the numerical simulations where sources have the
finite size.  \citet{keeton01a} only considered $\epsilon_{\rm th}=10$
case, but we also consider $\epsilon_{\rm th}=7$ and $4$ and find that
arcs with the smaller threshold values are more sensitively affected by
the source ellipticities. In
conclusion, except for the $\epsilon_{\rm th}=4$ case, the number ratio
of radial to tangential arcs seems not to be severely affected by source
ellipticity; the difference between
$\alpha=1.5$ and $\alpha=1.0$ is much larger than the change due to
the source ellipticity. 

\subsection{Lens Ellipticity}

\begin{figure*}
\begin{center}
\leavevmode
\epsfxsize=10cm 
\epsfbox{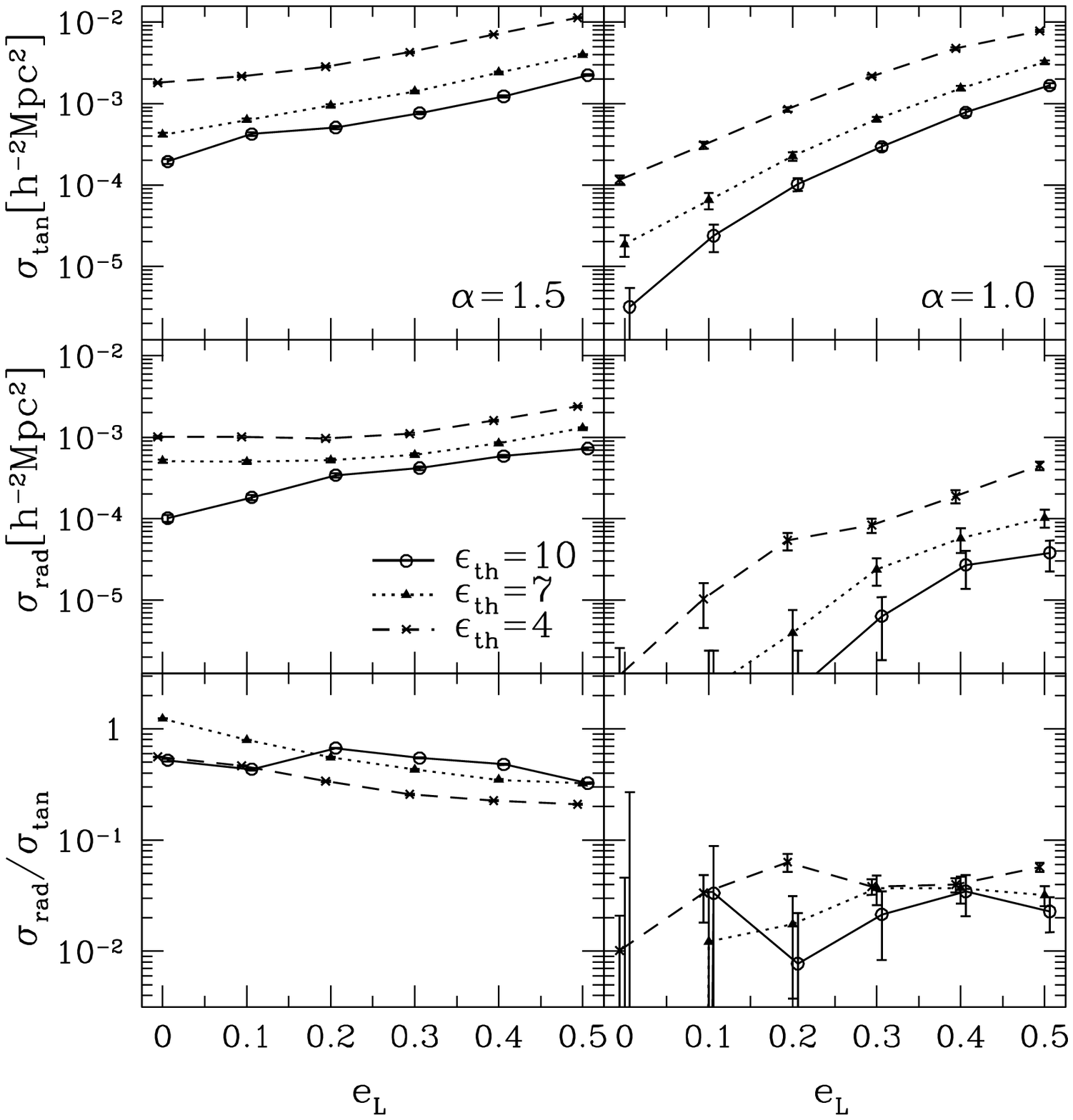} 
\caption{Same as Figure \ref{fig:sel}, but the plots against the lens
 ellipticity.\label{fig:el}}
\end{center}
\end{figure*}

The effect of the lens ellipticity is shown in Figure \ref{fig:el}. As
previously indicated by \citet{bartelmann95a}, lens ellipticity
drastically changes the total number of arcs; more than two orders of
magnitude from $e_{\rm L}=0$ to $e_{\rm L}=0.5$. The lens ellipticity
amplifies the numbers of both tangential and radial 
arcs. In particular, $\alpha=1.5$ profile and $\alpha=1.0$ profile can
produce almost the same number of tangential arcs when $e_{\rm L}=0.5$.
Nevertheless, the number ratio of radial to tangential arcs is not
affected much by the lens ellipticity. Even in the case of $e_{\rm
L}=0.5$, however, the number ratio of radial to tangential arcs changes
less than one order of magnitude compared with the spherical lens and is
still about one order of magnitude different between $\alpha=1.5$ and
$\alpha=1.0$. Therefore, we conclude that the number ratio remains a
good indicator for halo density profile which does not so affected by
the uncertainty of the lens ellipticity as to hide the difference caused
by the density profile.   

\subsection{Mass Dependence of Ellipticity Effects}

\begin{figure*}
\begin{center}
\leavevmode
\epsfxsize=10cm 
\epsfbox{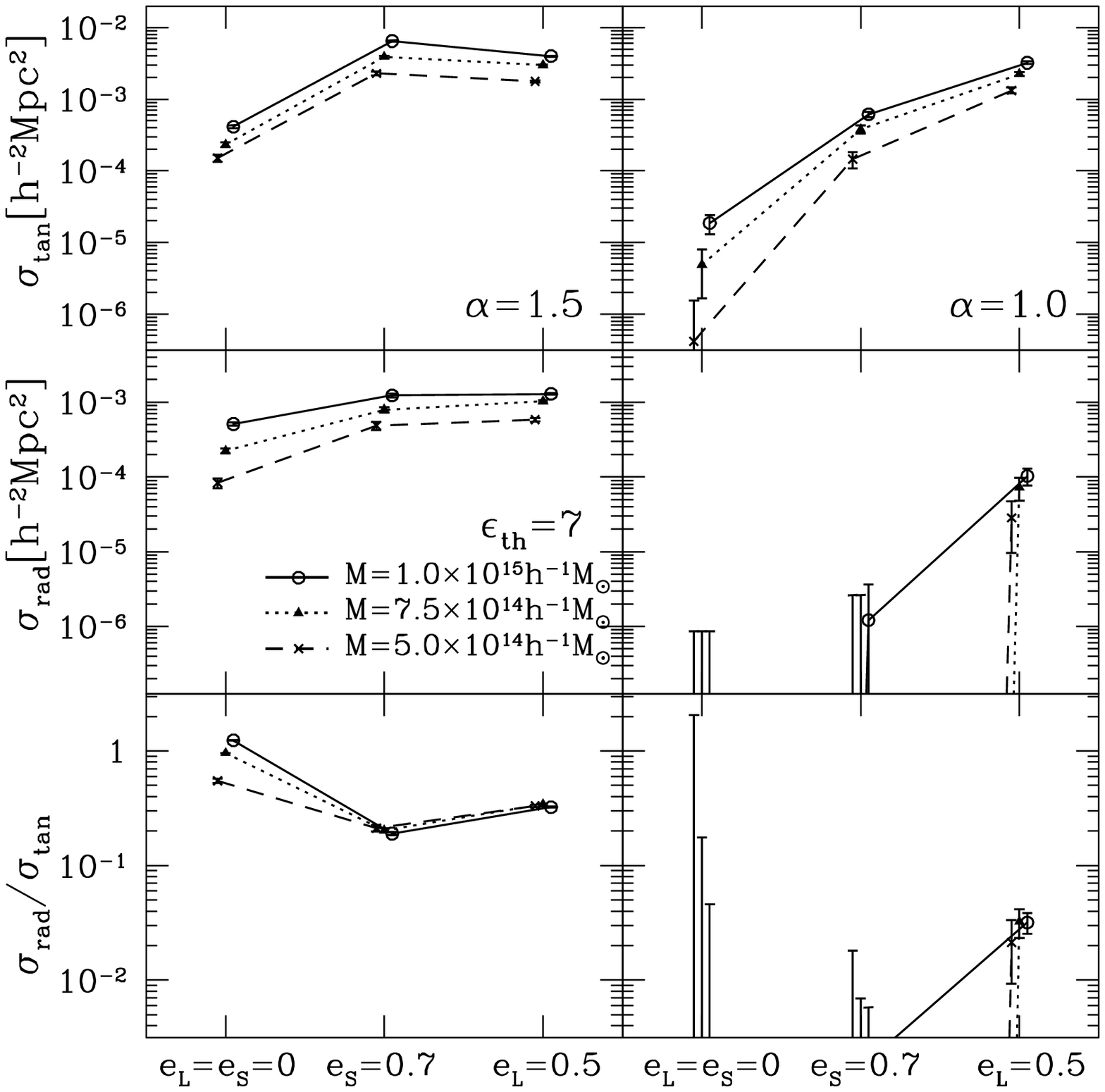} 
\caption{The effect of source and lens ellipticities for different lens
 mass. The cross sections for three halos with different mass are
 plotted; $M=1.0\times 10^{15}h^{-1}M_\odot$ ({\it solid}), 
$7.5\times 10^{14}h^{-1}M_\odot$ ({\it dotted}), and
$5.0\times 10^{14}h^{-1}M_\odot$ ({\it dashed}). In this plot we fix
 the threshold axis ratio as $\epsilon_{\rm th}=7$. For the effect of
 ellipticities, we examine only following three extreme cases; no
 ellipticities, $e_{\rm S}=0.7$, and $e_{\rm L}=0.5$.\label{fig:mass}}
\end{center}
\end{figure*}

In the above examples, we consider the massive halo with mass $M=1.0\times
10^{15}h^{-1}M_\odot$ only. While this is quite typical mass of lensing 
clusters, it is important to check whether our qualitative results 
described above remain valid for less massive halos. Therefore in this 
subsection we see the effects of source and lens ellipticities for halos
with different mass.

The results are shown in Figure \ref{fig:mass}. We plot the cross
sections for three halos with different mass; $M=1.0\times
10^{15}h^{-1}M_\odot$, $7.5\times 10^{14}h^{-1}M_\odot$, and
$5.0\times 10^{14}h^{-1}M_\odot$. In this plot we fix the threshold
axis ratio as $\epsilon_{\rm th}=7$ and examine only extreme cases,
$e_{\rm S}=0.7$ and $e_{\rm L}=0.5$, as well as the case of no
ellipticities. This figure clearly shows that our qualitative results
are not affected at all by changing the mass of the lens object; the
values of cross sections show quite similar behavior for halos with
different mass. Therefore our results are generic and applicable to a
wide range of the lens mass. Moreover, it seems that the number ratio of
radial to tangential arcs is highly insensitive to the mass of halos,
especially for the cases including ellipticities. This means that the
uncertainty arising from the lens mass in determining the density
profile is quite small.

\section{Validity of the Selection Criterion for Tangential and 
Radial Arcs}\label{sec:select}

\begin{figure*}
\begin{center}
\leavevmode
\epsfxsize=10cm 
\epsfbox{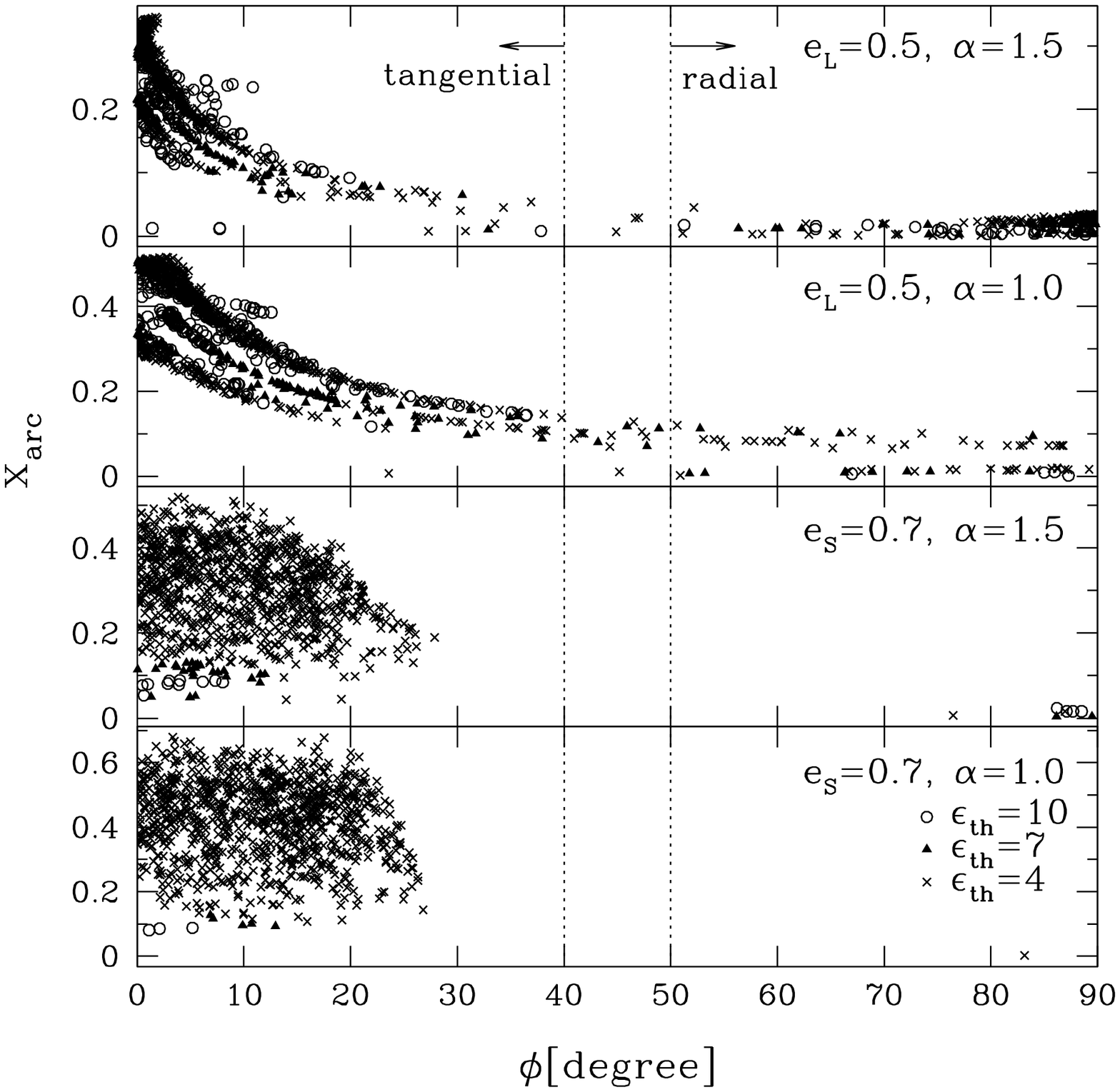} 
\caption{Plots for the arc positions $x_{\rm arc}$ against the
 orientations of arcs $\phi$ in the extreme cases we examined: 
 $e_{\rm L}=0.5$ and $e_{\rm S}=0.7$. In each plot, we only put the points 
 of randomly chosen 1000 arcs. Arcs with $\epsilon_{\rm th}=10$,
 $7$, and $4$ are shown by open circles, filled triangles, and crosses,
 respectively. As shown in equations (\ref{detecttan}) and 
 (\ref{detectrad}), we regard arcs with $\phi\leq40^\circ$ as tangential arcs 
 and with $\phi\geq50^\circ$ as radial arcs.\label{fig:incpos}}
\end{center}
\end{figure*}

In equations (\ref{detecttan}) and (\ref{detectrad}), we discriminated 
tangential and radial arcs by the orientation $\phi$. We regard arcs with 
$\phi\leq40^\circ$ as tangential arcs and with $\phi\geq50^\circ$ as 
radial arcs. There is, however, no obvious reason to adopt this selection 
criterion. It may be possible that tangential and radial arcs are 
misinterpreted with each other. Therefore, we check the validity of 
our selection criterion in the extreme cases we examined: $e_{\rm
L}=0.5$ and $e_{\rm S}=0.7$. Figure \ref{fig:incpos} shows the distances
of arcs from the center of the lens against the orientations of arcs
defined in \S \ref{sec:rec}. This figure suggests that our selection
criterion distinguishes tangential and radial arcs quite well for
$e_{\rm S}=0.7$ cases. In the $e_{\rm L}=0.5$ cases, however, there
seems to be mixing between both arcs. In particular, the case with
$e_{\rm L}=0.5$ and $\alpha=1.0$ produces ``radial'' arcs which have
$\phi\geq 50^\circ$ but seem to be connected with tangential arcs in the
$\phi$-$x_{\rm arc}$ plane. The fraction of such arcs, however, is small
for longer arcs ($\epsilon_{\rm th}\gtrsim 7$). Therefore, it is safe to
use our selection criterion of tangential (eq. [\ref{detecttan}]) and
radial arcs (eq. [\ref{detectrad}]), especially for $\epsilon_{\rm
th}\gtrsim 7$. 

\section{Summary}\label{sec:summary}
We have studied several systematic effects which may change the number of 
tangential and radial arcs. More specifically, we have examined the effect 
of the finite source size, source ellipticity, and lens ellipticity. 

First we derived an expression of the cross section for radial arcs
(eq. [\ref{cs_rad}]) 
taking account of the finite source size, assuming spherical lenses. 
We found that the cross section is enhanced by the moderate amount of
the source size, while larger source size (source diameter $\gtrsim
1''$) rapidly decreases the probability of producing the radial arcs. We
also derived the cross section for tangential and radial arcs
numerically and confirmed that radial arcs are more sensitive to the
finite source size effect than tangential arcs. It has been also found
that our analytic prediction of the cross section for radial arcs shows
fairly good agreement with the numerical simulations. Therefore, one can
accurately predict the number of radial arcs with an arbitrary source
size. 

In addition to the finite source size effect, we studied effects of 
lens and source ellipticities. We found that both ellipticities can 
change the number of both tangential and radial arcs significantly; the 
enhancement can become as large as two orders of magnitude. The number
ratio of radial to tangential arcs  is, however, not so affected by
ellipticities. Even in the extremely elliptical cases, the number ratio
changes less than one order of magnitude on average, and still remains
the difference between $\alpha=1.5$ and $\alpha=1$ profiles. The
exception is the arcs with $\epsilon_{\rm th}=4$. Such short arcs are
more severely affected by the source ellipticity; the number ratio
becomes much smaller compared with no ellipticity case. Although the
these results are obtained for one specific halo with mass
$M=10^{15}h^{-1}M_\odot$, we also examined the effects of ellipticities
for several halos with different mass and confirmed that our qualitative
results remain valid. As a conclusion, number ratio of radial and
tangential arcs with high threshold axis ratio $\epsilon_{\rm
th}\gtrsim7$ is quite robust for the uncertainties of ellipticities. 

In summary, the number ratio of radial to tangential arcs which have the
large axis ratio  ($\gtrsim 7$) becomes good statistics which can probe
the density profile of the lens object, if the source size effect is
correctly taken into account. If we use the maximum cross section
for radial arcs for the theoretical prediction, observations of
tangential and radial arcs mainly give the lower limit of the inner
slope $\alpha$ because the effects of ellipticities as well as the
source size decrease the fraction of radial arcs. On the other hand, the
total number of tangential arcs, which is also highly sensitive to the
density profile of the lens object \citep{oguri01}, should be carefully
calibrated with the ellipticities of the lens and source. Since the
detailed non-spherical modeling of dark halos is now being attempted
\citep{jing02}, it would become an important task to take account of the
effect of the lens ellipticity systematically.

Of course, there are other effects of simplifications we should
consider. First, we neglect the irregularity of mass distribution in
clusters. This effect, however, seems to be small enough
\citep{flores00,meneghetti00}. Secondly, we neglect the effect of
central cD galaxies. Radial arcs occur very close to cluster center,
thus it is possible that radial arcs are hidden by the light of cD
galaxies. Multicolor imaging, however, may help the identification of
radial arcs because arcs have rather different color compared with those
of member galaxies \citep{molikawa01}. More importantly, the
gravitational potential of cD galaxies may have serious effects on arcs,
especially for radial arcs \citep{miralda95,williams99,molikawa01}.
Discriminating the dark halo profile from the cD galaxy profile in the
central region is in general difficult because the gravitational lensing
can probe only the sum of these profiles. Therefore, it may be needed to
use the sample of clusters without cD galaxies to probe the density
profile of dark halos.   

\acknowledgments
I would like to thank Yasushi Suto and Atsushi Taruya for useful
discussions and comments. I also thank an anonymous referee for helpful
comments.



\begin{thebibliography}{}
\bibitem[Bartelmann(1995)]{bartelmann95b}
Bartelmann, M. 1995, \aap, 299, 11
\bibitem[Bartelmann(1996)]{bartelmann96}
Bartelmann, M. 1996, \aap, 313, 697
\bibitem[Bartelmann et al.(1995)Bartelmann, Steinmetz, \& Weiss]{bartelmann95a}
Bartelmann, M., Steinmetz, M., \& Weiss, A. 1995, \aap, 297, 1
\bibitem[Bartelmann \& Weiss(1994)]{bartelmann94}
Bartelmann, M., \& Weiss, A. 1994, \aap, 287, 1
\bibitem[Bullock et al.(2001)]{bullock01}
Bullock, J.~S., Kolatt, T.~S., Sigad, Y., Somerville, R.~S., 
Kravtsov, A.~V., Klypin, A.~A., Primack, J.~R., \& Dekel, A. 2001, 
\mnras, 321, 559
\bibitem[Flores et al.(2000)Flores, Maller, \& Primack]{flores00}
Flores, R.~A., Maller, A.~H., \& Primack, J.~R. 2000, \apj, 535, 555
\bibitem[Golse \& Kneib(2002)]{golse02}
Golse, G., \& Kneib, J.~P. 2002, \aap, submitted (astro-ph/0112138)
\bibitem[Hattori et al.(1997)Hattori, Watanabe, \& Yamashita]{hattori97}
Hattori, M., Watanabe, K., \& Yamashita, K. 1997, \aap, 319, 764
\bibitem[Jing \& Suto(2000)]{jing00}
Jing, Y.~P., \& Suto, Y. 2000, \apjl, 529, L69
\bibitem[Jing \& Suto(2002)]{jing02}
Jing, Y.~P., \& Suto, Y. 2002, \apj, in press (astro-ph/0202064)
\bibitem[Keeton(2001)]{keeton01a}
Keeton, C.~R. 2001, \apj, 562, 160
\bibitem[Keeton \& Madau(2001)]{keeton01b}
Keeton, C.~R., \& Madau, P. 2001, \apjl, 549, L25
\bibitem[Lilly, Cowie, \& Gardner(1991)]{lilly91}
Lilly, S.~J., Cowie, L.~L., \& Gardner, J.~P. 1991, \apj, 369, 79
\bibitem[Meneghetti et al.(2000)]{meneghetti00}
Meneghetti, M., Bolzonella, M., Bartelmann, M., 
Moscardini, L., \& Tormen, G. 2000, \mnras, 314, 338
\bibitem[Meneghetti et al.(2001)]{meneghetti01}
Meneghetti, M., Yoshida, N., Bartelmann, M., Moscardini, L., 
Springel, V., Tormen, G., \& White S.~D.~M. 2001, 
\mnras, 325, 435
\bibitem[Meneghetti et al.(2002)Meneghetti, Bartelmann, \& Moscardini]{meneghetti02}
Meneghetti, M., Bartelmann, M., \& Moscardini, L. 2002, \mnras,
submitted (astro-ph/0201501)
\bibitem[Miralda-Escud\'{e}(1993a)]{miralda93a}
Miralda-Escud\'{e}, J. 1993a, \apj, 403, 497
\bibitem[Miralda-Escud\'{e}(1993b)]{miralda93b}
Miralda-Escud\'{e}, J. 1993b, \apj, 403, 509
\bibitem[Miralda-Escud\'{e}(1995)]{miralda95}
Miralda-Escud\'{e}, J. 1995, \apj, 438, 514
\bibitem[Molikawa \& Hattori(2001)]{molikawa01}
Molikawa, K., \& Hattori, M. 2001, \apj, 559, 544
\bibitem[Molikawa et al.(1999)]{molikawa99}
Molikawa, K., Hattori, M., Kneib, J.~P., \& Yamashita, K. 1999, 
\aap, 351, 413
\bibitem[Navarro et al.(1996)Navarro, Frenk, \& White]{navarro96}
Navarro, J.~F., Frenk, C.~S., \& White, S.~D.~M. 1996, \apj, 462, 563
\bibitem[Navarro et al.(1997)Navarro, Frenk, \& White]{navarro97}
Navarro, J.~F., Frenk, C.~S., \& White, S.~D.~M. 1997, \apj, 490, 493
\bibitem[Oguri et al.(2001)Oguri, Taruya, \& Suto]{oguri01}
Oguri, M., Taruya, A., \& Suto, Y. 2001, \apj, 559, 572
\bibitem[Schneider et al.(1992)Schneider, Ehlers, \& Falco]{schneider92}
Schneider, P., Ehlers, J., \& Falco, E. E. 1992, 
Gravitational Lenses (New York: Springer)
\bibitem[Tyson et al.(1998)Tyson, Kochanski, \& Dell'Antonio]{tyson98}
Tyson, J.~A., Kochanski, G.~P., \& Dell'Antonio I.~P. 1998, \apjl, 498, L107 
\bibitem[William et al.(1999)Williams, Navarro, \& Bartelmann]{williams99}
Williams, L.~L.~R., Navarro, J.~F., \& Bartelmann, M. 1999, \apj, 527, 535
\bibitem[Wu \& Hammer(1993)]{wu93}
Wu, X.-P., \& Hammer, F. 1993, \mnras, 262, 187
\end{thebibliography}
\end{document}